\documentclass[prb,twocolumn,superscriptaddress,showpacs]{revtex4}

\usepackage{graphicx}% Include figure files
\usepackage{dcolumn}% Align table columns on decimal point
\usepackage{bm}% bold math
\usepackage{tabularx}% Table width
\usepackage{amsmath}% Higher math

\usepackage{amssymb}
\usepackage{txfonts}
\usepackage{graphics,color,epsfig}
\usepackage{epstopdf}

\begin{document}

\title{Universal critical behavior in the ferromagnetic superconductor Eu(Fe$_{0.75}$Ru$_{0.25}$)$_{2}$As$_{2}$}

\author{Zheng Zhou}
\affiliation{State Key Laboratory of Surface Physics and Department of Physics, Fudan University, Shanghai 200433, China}

\author{W. T. Jin}
\email{jwt2006@gmail.com}
\affiliation{J\"{u}lich Centre for Neutron Science (JCNS) at Heinz Maier-Leibnitz Zentrum (MLZ), Forschungszentrum J\"{u}lich GmbH, Lichtenbergstra{\ss}e 1, D-85747 Garching, Germany}

\author{Wei Li}
\email{w_li@fudan.edu.cn}
\affiliation{State Key Laboratory of Surface Physics and Department of Physics, Fudan University, Shanghai 200433, China}
\affiliation{Collaborative Innovation Center of Advanced Microstructures, Nanjing 210093, China}

\author{S. Nandi}
\affiliation{J\"{u}lich Centre for Neutron Science (JCNS) and Peter Gr\"{u}nberg Institut PGI, JARA-FIT, Forschungszentrum J\"{u}lich GmbH, D-52425 J\"{u}lich, Germany}

\author{B. Ouladdiaf}
\affiliation{Institut Laue Langevin, 71 rue des Martyrs, Bo\^{\i}te Postale 156, 38042 Grenoble Cedex 9, France}

\author{Zheng Yan}
\email{zhengyan13@fudan.edu.cn}
\affiliation{State Key Laboratory of Surface Physics and Department of Physics, Fudan University, Shanghai 200433, China}

\author{Xinyuan Wei}
\affiliation{State Key Laboratory of Surface Physics and Department of Physics, Fudan University, Shanghai 200433, China}

\author{Xuguang Xu}
\affiliation{School of Physical Science and Technology, ShanghaiTech University, Shanghai 201210, China}

\author{W. H. Jiao}
\affiliation{School of Science, Zhejiang University of Science and Technology, Hangzhou 310023, China}

\author{N. Qureshi}
\affiliation{Institut Laue Langevin, 71 rue des Martyrs, Bo\^{\i}te Postale 156, 38042 Grenoble Cedex 9, France}

\author{Y. Xiao}
\affiliation{School of Advanced Materials, Peking University Shenzhen Graduate School, Shenzhen 518055, China}

\author{Y. Su}
\affiliation{J\"{u}lich Centre for Neutron Science (JCNS) at Heinz Maier-Leibnitz Zentrum (MLZ), Forschungszentrum J\"{u}lich GmbH, Lichtenbergstra{\ss}e 1, D-85747 Garching, Germany}

\author{G. H. Cao}
\affiliation{Department of Physics, Zhejiang University, Hangzhou 310027, China}

\author{Th. Br\"{u}ckel}
\affiliation{J\"{u}lich Centre for Neutron Science (JCNS) and Peter Gr\"{u}nberg Institut PGI, JARA-FIT, Forschungszentrum J\"{u}lich GmbH, D-52425 J\"{u}lich, Germany}
\affiliation{J\"{u}lich Centre for Neutron Science (JCNS) at Heinz Maier-Leibnitz Zentrum (MLZ), Forschungszentrum J\"{u}lich GmbH, Lichtenbergstra{\ss}e 1, D-85747 Garching, Germany}

\date{\today}

\begin{abstract}
The study of universal critical behavior is a crucial issue in a continuous phase transition, which groups various critical phenomena into universality classes for revealing microscopic electronic behaviors. The understanding of the nature of magnetism in Eu-based ferromagnetic superconductors is largely impeded by the infeasibility of performing inelastic neutron scattering measurements to deduce the microscopic magnetic behaviors and the effects on the superconductivity, due to the significant neutron absorption effect of
natural $^{152}$Eu and unavailability of large single crystals. However, by systematically combining the neutron diffraction experiment, the first-principles calculations, and the quantum Monte Carlo simulations, we have obtained a perfectly consistent universal critical exponent value of $\beta=0.385(13)$ experimentally and theoretically for Eu(Fe$_{0.75}$Ru$_{0.25}$)$_{2}$As$_{2}$, from which the magnetism in the Eu-based ferromagnetic superconductors is identified as the universal class of a three-dimensional anisotropic quantum Heisenberg model with long-range magnetic exchange coupling. This study not only clarifies the nature of microscopic magnetic behaviors in the Eu-based ferromagnetic superconductors, but also opens a new avenue of systemic methodology for studying the universal critical behaviors associated with magnetic phase transitions in the area of magnetism and the spin fluctuations effects on the unconventional superconductivity.
\end{abstract}

\pacs{}
\maketitle

%\section{Introduction}
{\it Introduction.---}The continuous phase transition associated symmetry breaking is one of the two central themes in Landau's theory in condensed matter physics~\cite{Landau}. An order parameter is introduced to well describe the symmetry changing across the boundaries in a phase transition. For instance, the order parameter is the net magnetization in a ferromagnetic system or the energy gap for Cooper pairs' formation in a superconductor undergoing a phase transition. Furthermore, the universal critical exponents and scaling functions are used to describe the behavior of physical quantities near continuous phase transitions, which are independent of the microscopic details of the systems, but only of some of their global properties, such as the space dimensionality, the range of interaction, and the symmetry of the order parameter. Unveiling the universal critical exponents of phase transitions in unconventional superconductors associated with the magnetic phase transition may shed light on the study of spin fluctuations effects on the superconducting mechanism.

\begin{figure*}
\centering{}\includegraphics[bb=10 10 800 300,width=16cm,height=6cm]{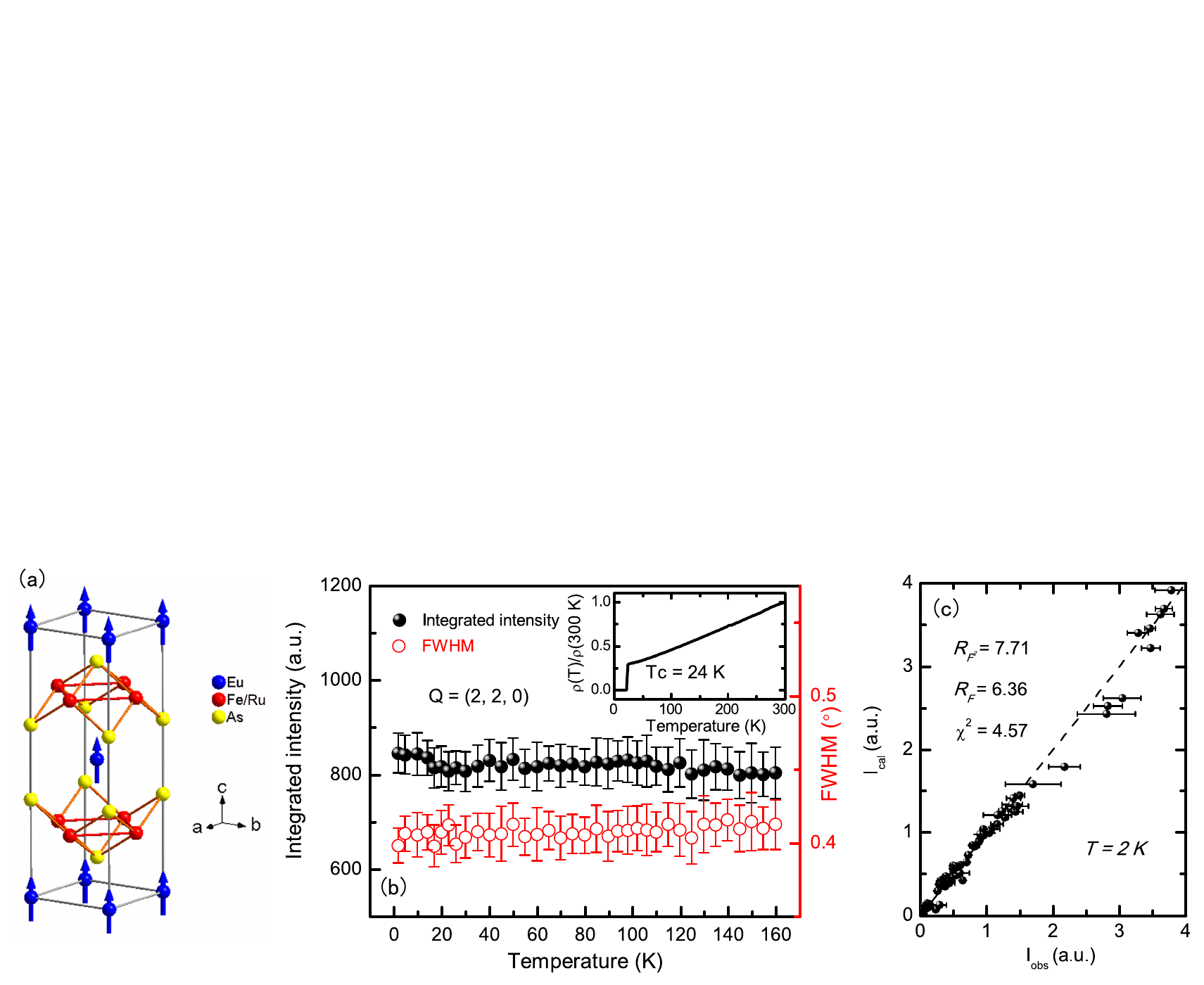}
\caption{(Color online) (a) The illustration of crystal and magnetic structure for Eu(Fe$_{0.75}$Ru$_{0.25}$)$_{2}$As$_{2}$. (b) The temperature dependence of the integrated intensity (solid spheres) and peak width (FWHM, open circles) of the (2, 2, 0)$_{T}$ nuclear reflection, respectively. Note that the slight increase of integrated intensity below 20 K is due to the ferromagnetic ordering of Eu. The inset shows the normalized in-plane resistivity of the Eu(Fe$_{0.75}$Ru$_{0.25}$)$_{2}$As$_{2}$ single crystal, in which no anomaly due to the structure phase transition and spin-density wave order is observed. (c) Comparison between the observed and calculated integrated intensities of the unique reflections at 2 K.}
\label{fig1}
\end{figure*}

In general, the ferromagnetism is incompatible with superconductivity for singlet pairing superconductors, since the superconductivity is suppressed or quenched whenever ferromagnetism appears~\cite{Bulaevskii}. The intriguing coexistence of superconductivity and ferromagnetism in the Eu-based iron pnictides upon chemical doping or applying external
pressure attracted enormous attentions~\cite{HSJeevan,ZRen,GCao,SJiang,WTJin2017}. One scenario to reach the compromise between the two antagonistic phenomena is the formation of a spontaneous vortex state without applying an external magnetic field~\cite{WHJiao}. Neutron diffraction experimentally confirmed the bulk nature of the ferromagnetism from Eu $4f$ orbitals with an ordered moment of $\sim7\mu_B$ per Eu atom and well suppressed antiferromagnetism of Fe $3d$ orbitals associated with the bulk superconductivity~\cite{WTJin2013,SNandi2014,WTJin2015}. Theoretically, the indirect Ruderman-Kittel-Kasuya-Yosida (RKKY) interaction via the mediated itinerant electrons on Fe $3d$ orbitals is proposed to be responsible for the ferromagnetism of the Eu sublattice~\cite{WeiLi,Akbari}. However, due to the infeasibility of performing inelastic neutron scattering measurements on Eu-rich materials with significant neutron absorptions of natural $^{152}$Eu, the microscopic magnetic exchange couplings cannot be determined experimentally, which largely impeded the thorough understanding of the magnetic behaviors of Eu $4f$ electrons in these ferromagnetic superconductors. Fortunately, the universal critical exponents provide an alternative to approach the nature of ordered magnetism. Available experimental studies on Eu-based pnictides have obtained the same critical exponent $\beta = 0.35$ for EuFe$_2$As$_2$~\cite{JKoo} and EuNi$_2$As$_2$~\cite{Jin2019}, fitting well into the universal class of three-dimensional isotropic quantum Heisenberg model class~\cite{SM}. In contrast, $\beta=0.32$ was obtained for EuRh$_2$As$_2$~\cite{SNandi2009}, more consistent with a three-dimensional Ising model ($\beta = 0.326$)~\cite{SM,Kagawa}. Furthermore, to the best of our knowledge, no studies on the critical behaviors of Eu magnetism in the doping induced ferromagnetic superconductors were performed yet, as the mostly used two methods to extract the critical exponents, either magnetometry or calorimetry, do not apply due to the interference of superconductivity.

In this Rapid Communication, we have developed a systematic method of combining the neutron diffraction measurements, the first-principles calculations, and the quantum Monte Carlo simulations to study the universal critical behaviors associated with a magnetic phase transition. A neutron diffraction experiment on Eu(Fe$_{0.75}$Ru$_{0.25}$)$_{2}$As$_{2}$, an isovalent doped ferromagnetic superconductor, found the critical exponent $\beta=0.386$. Based on the first-principles calculations of its electronic structure, we noticed that the occupied states of Eu $4f$ orbitals located well below the Fermi level, indicating the suitability of a localized anisotropic Heisenberg model for describing the Eu-magnetism. Further applying a quantum Monte Carlo algorithm and using the computed values of magnetic exchange coupling, $\beta=0.386$ was found theoretically, in excellent agreement with the value extracted experimentally. The magnetism in the studied Eu-based ferromagnetic superconductor is thus identified as the universal class of three-dimensional anisotropic quantum Heisenberg model with long-range magnetic exchange coupling. This finding points out a suitable microscopic theoretical model for describing the nature of magnetism in the intriguing Eu-based ferromagnetic superconductors.

%The rest of this paper is organized as follows. In Sec.~\ref{Sec2}, we first introduce the universal critical exponent result evaluated from the neutron scattering experiments. The first-principles calculations are carried out in Sec.~\ref{Sec3} to establish a microscopic theoretical model for describing the ferromagnetism. In Sec.~\ref{Sec4}, the theoretical universal exponent critical value is evaluated from the quantum Monte Carlo calculations based on the nature of first-principles calculations. Finally, Sec.~\ref{final} is devoted to a brief discussion and its summary.

%\section{Neutron Scattering Experiments}\label{Sec2}

{\it Experimental results.---}Single crystals of the Eu-based material Eu(Fe$_{1-x}$Ru$_{x}$)$_{2}$As$_{2}$ ($x$ = 0.25) were grown from self-flux (Fe, Ru)As, and well characterized to be a ferromagnetic superconductor by electric resistivity, magnetization and M\"{o}ssbauer measurements~\cite{Jiao_11}. A single crystal with the mass $\sim5$ mg and dimensions $\sim3$ $\times$ 2 $\times$ 0.2 mm$^{3}$ from the same batch was selected for the neutron diffraction experiment (see the Supplemental Material~\cite{SM} for the experimental details), which was performed on the four-circle thermal-neutron diffractometer D10 at the Institut LaueLangevin (Grenoble, France). The crystal and magnetic structure of isovalent ruthenium doped Eu(Fe$_{0.75}$Ru$_{0.25}$)$_{2}$As$_{2}$ at 2 K determined by neutron diffraction is illustrated in Fig.~\ref{fig1}(a). Since the parent compound EuFe$_2$As$_2$ shows a spin-density wave (SDW) order in the Fe sublattice accompanied by a structure phase transition at 190 K~\cite{Xiao_09}, we have tracked a strong nuclear reflection upon cooling, which is sensitive to the tetragonal-orthorhombic structure phase transition signaled by a sudden increase of its intensity and broadening of its width~\cite{WTJin2013, WTJin2017}. Fig.~\ref{fig1}(b) displays the temperature dependence of the integrated intensity and the full width at half maximum (FWHM) of the (2, 2, 0)$_{T}$ peak in the tetragonal notation, respectively. It is shown that both of them keep almost constant, indicating the structure phase transition is fully suppressed by 25 \% isovalent Ru doping. In addition, no intensities were observed at the $\mathbf{Q}$ points with propagation vector $\vec{k}$ = (1/2, 1/2, 1)$_{T}$ even with the analyzer, excluding any residual Fe SDW order in this bulk superconductor. The absence of both structure phase transition and SDW order in Eu(Fe$_{0.75}$Ru$_{0.25}$)$_{2}$As$_{2}$ is consistent with the temperature dependence of its in-plane electric resistivity shown in the inset of Fig.~\ref{fig1}(b), in which no anomaly in addition to the sharp superconducting transition at 24 K is observed.

The integrated intensities of 241 reflections at 30 K and 304 reflections at 2 K were collected, respectively, to determine the details of the ground-state ferromagnetic structure of Eu(Fe$_{0.75}$Ru$_{0.25}$)$_{2}$As$_{2}$. After necessary absorption corrections using the DATAP program~\cite{Coppens_65}, the equivalent reflections were merged into the unique ones based on the tetragonal symmetry. At 30 K, which is above the magnetic ordering temperature of Eu, the nuclear structure is refined using the FULLPROF program within the $\mathit{I4/mmm}$ space group~\cite{Rodriguez-Carvajal_93}. The occupancy of Ru was refined to be 22(5)\%, consistent with the value of 25\% determined from the energy dispersive x-ray spectroscopy~\cite{Jiao_11}. At 2 K, additional intensities appear on top of the nuclear reflections measured at 30 K, suggesting a magnetic propagation vector of $\vec{k}$ = 0 for the Eu sublattice. According to the irreducible representation analysis~\cite{Wills_00}, only ferromagnetic structures with the moments aligned along the $\mathit{c}$-axis or in the $\mathit{ab}$ plane are allowed for the Eu$^{2+}$ spins due to symmetry restriction. However, invariant integrated intensities of the (0, 0, even) reflections at 2 K and 30 K exclude the possibility of in-plane ferromagnetic alignment. As shown in Fig.~\ref{fig1}(c), adding a ferromagnetic Eu$^{2+}$ moment of 7.0(2) $\mathit{\mu_{B}}$ along the $\mathit{c}$-axis into the nuclear structure determined from 30 K yields a rather good fitting to the intensities at 2 K. The parameters of the nuclear and magnetic structures of Eu(Fe$_{0.75}$Ru$_{0.25}$)$_{2}$As$_{2}$ at 2 K determined by the refinements are given in Table~\ref{tabel_1}.

\begin{table}[bp]
\caption{Parameters of the nuclear and magnetic structures of Eu(Fe$_{0.75}$Ru$_{0.25}$)$_{2}$As$_{2}$ at 2 K obtained from refinements of single-crystal neutron diffraction data. The occupancy of Ru was refined to be 22(5)\%. [Space group: $\mathit{I4/mmm}$, $\mathit{a}$ = 3.953(3)\AA, $\mathit{c}$ = 11.567(4)\AA]}
\begin{ruledtabular} %
\begin{tabular}{ccccc}
Atom/site & $\mathit{x}$ & $\mathit{y}$ & $\mathsf{\mathit{z}}$ & $B\,$(\AA\textsuperscript{2})\tabularnewline
\hline
Eu ($2a$) & 0 & 0 & 0 & 0.07(4)\tabularnewline
\multicolumn{5}{c}{$M_{Eu}$= ($\mathit{M_{a}},\mathit{M_{b}},\mathit{M_{c}})$ = (0, 0, 7.0(2)) $\mu_{B}$}\tabularnewline
Fe/Ru ($4d$) & 0.5 & 0 & 0.25 & 0.10(4)\tabularnewline
As ($4e$) & 0 & 0 & 0.3617(3) & 0.13(5)\tabularnewline
\end{tabular}\end{ruledtabular}
\label{tabel_1}
\end{table}

\begin{figure}
\centering{}\includegraphics[bb=15 850 550 1140,width=8cm,height=4.5cm]{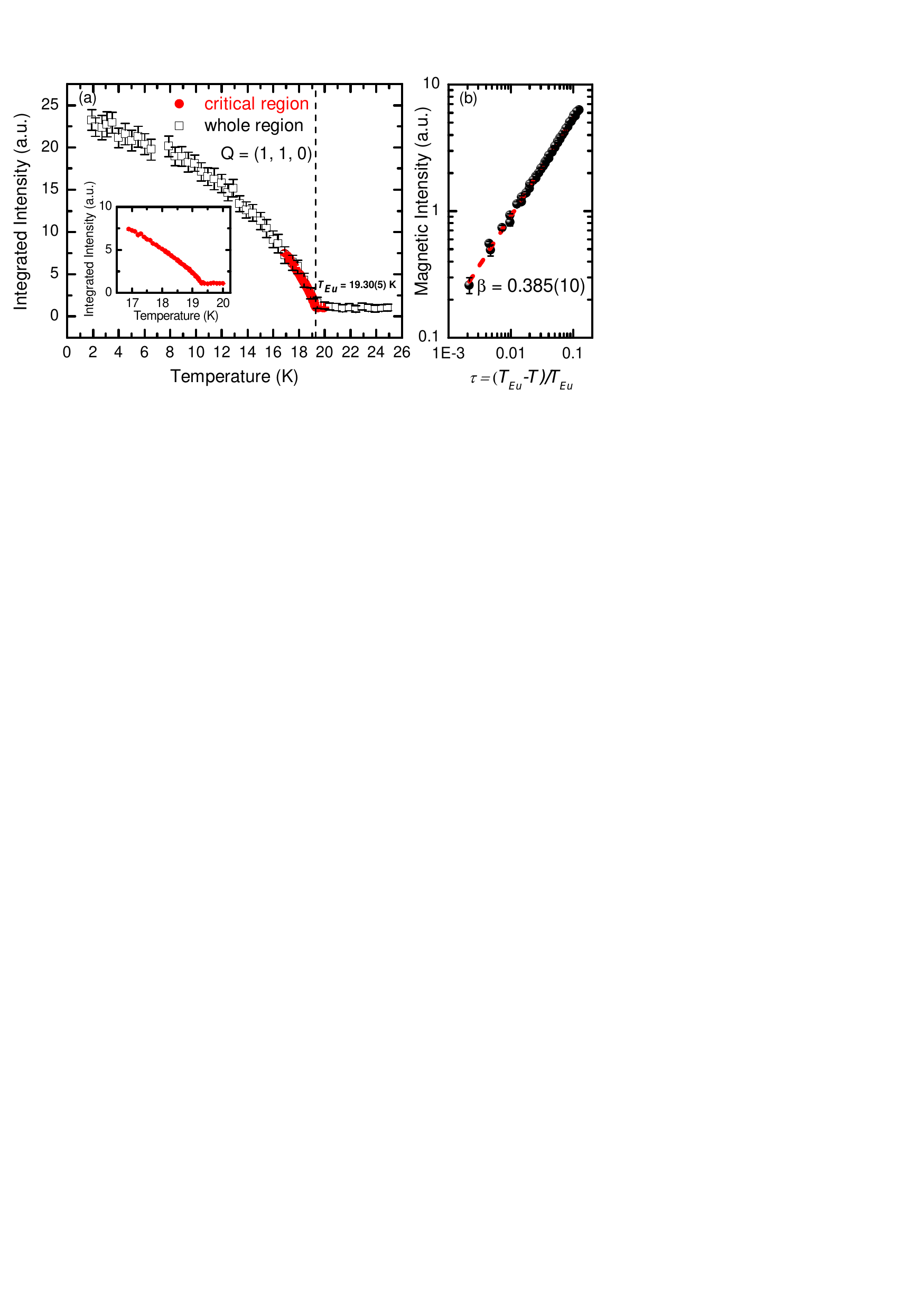}
\caption{(Color online) The temperature dependence of the integrated intensity of the (1, 1, 0) reflection in the whole temperature region (open squares) and critical region [filled circles, also shown in the inset of (a) as enlarged plot], respectively. The vertical dashed line marks the ferromagnetic ordering temperature of the Eu$^{2+}$ moments. The linear fitting (dashed line) to the magnetic intensity, $\mathit{I_{M}(\tau)}$, in the double logarithmic plot is shown in (b).}
\label{fig2}
\end{figure}

Fig.~\ref{fig2} shows the temperature dependence of the integrated intensity of the (1, 1, 0) reflection, from which the ferromagnetic ordering temperature of the Eu$^{2+}$ moments can be determined to be $\mathit{T_{Eu}}$= 19.30(5) K. The excellent stability and accuracy of temperature control within 0.05 K at the D10 diffractometer provides the unique chance to investigate the critical behavior close to the ferromagnetic transition of Eu. Although we can not observe the magnetic diffuse scattering due to spin fluctuations above the transition temperature with such a small single crystal, careful measurements of the integrated intensity of from 17 K to 20 K with small temperature steps [as shown in the inset of Fig.~\ref{fig2}(a)] allows us to extract the universal critical exponent $\beta$ of the ferromagnetic phase transition. After subtracting the nuclear contribution above $\mathit{T_{Eu}},$ the magnetic diffraction intensity can be fitted using the power law $\mathit{I_{M}}$ $\propto$ $\mathit{M^{2}}$ $\propto$ $\tau^{2\beta}$, where $\mathit{M}$ is the magnetic order parameter and $\tau$ = $\frac{T_{Eu}-T}{T_{Eu}}$. By linear fitting (dashed line) to $\mathit{I_{M}(\tau)}$ in the double logarithmic plot [Fig.~\ref{fig2}(b)], the universal critical exponent $\beta$ is deduced to be 0.385 $\pm$ 0.010, which is significantly larger than that of $\beta=0.35$ for EuFe$_2$As$_2$~\cite{JKoo}. Since the Ru 4$d$ orbitals are much more extended than the Fe 3$d$ orbitals, the RKKY-type long-range coupling between Eu atoms mediated by the Fe 3$d$ electrons on FeAs layers may get enhanced in Eu(Fe$_{1-x}$Ru$_{x}$)$_{2}$As$_{2}$ with Ru doping, resulting in the enhancement of long-range magnetic coupling in Eu layers. Furthermore, the three-dimensional isotropic quantum Heisenberg model displays a smaller critical exponent value of $\beta=0.365$~\cite{LZhang}. Thus, from the viewpoint of theory, we expect that the relatively large universal critical exponent $\beta=0.385(13)$ observed in the ferromagnetic superconductor Eu(Fe$_{0.75}$Ru$_{0.25}$)$_{2}$As$_{2}$ might suggest a strong anisotropy in a three-dimensional quantum Heisenberg model with long-range magnetic exchange coupling.

%\section{The First-Principles Calculations}\label{Sec3}

{\it First-principles calculations.---}Before evaluating numerically the universal critical exponent $\beta$ for the ferromagnetic superconductor Eu(Fe$_{0.75}$Ru$_{0.25}$)$_{2}$As$_{2}$, we have performed first-principles calculations to establish a microscopic theoretical model for describing the Eu ferromagnetism. The calculations were performed using the all-electron full potential linear augmented plane wave plus local orbitals (FP-LAPW+lo) method~\cite{DJSingh} as implemented in the WIEN2K code~\cite{PBlaha}. The exchange-correlation potential was calculated using the generalized gradient approximation (GGA) as proposed by Pedrew, Burke, and Ernzerhof~\cite{PBE,potentials}. We have included the strong Coulomb repulsion in the Eu 4$f$ orbitals on a mean-field level using the GGA $+U_{eff}$ approximation, applying the atomic limit double-counting scheme. Throughout this Rapid Communication, we have used a $U_{eff}$ of $8$ eV, which is the standard value for an Eu$^{2+}$ ion~\cite{WeiLi}, while we did not apply $U_{eff}$ to the itinerant Fe 3$d$ orbitals. The results were also checked for consistency with varying $U_{eff}$ values. In addition, the spin-orbit coupling was also included with the second variational method in the Eu 4$f$ orbitals.

\begin{figure}
\centering{}\includegraphics[width=8cm,height=6cm]{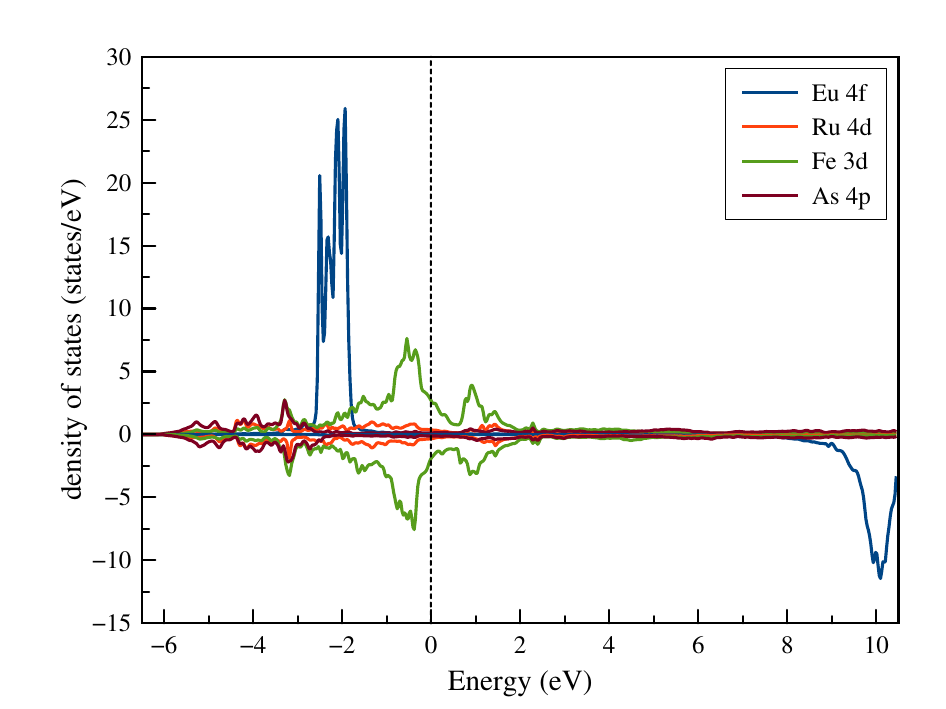}
\caption{(Color online) The calculated projected density of states on the Eu 4$f$, Ru 4$d$, Fe 3$d$, and As 4$p$ orbitals per unit cell of Eu(Fe$_{0.75}$Ru$_{0.25}$)$_{2}$As$_{2}$. The Fermi energy is set to zero (black dashed line).}
\label{fig3}
\end{figure}

The calculated projected density of states on the orbitals of Eu 4$f$, Ru 4$d$, Fe 3$d$ and As 4$p$ for Eu(Fe$_{0.75}$Ru$_{0.25}$)$_{2}$As$_{2}$ are shown in Fig.~\ref{fig3} based on the supercell method. Since the Eu 4$f$ orbitals are quite localized, the Eu ions are in a stable $2+$ valence state with a half-filled 4$f$ shell, resulting in the ferromagnetic order of Eu$^{2+}$ spins with the magnetic moment of 7 $\mu_B$. As can be seen clearly in Fig.~\ref{fig3}, the spin-up components of Eu 4$f$ states are located lower than -2 eV below the Fermi level, while the spin-down components are unoccupied and located well above the Fermi level (larger than 10 eV). Near the Fermi level, the main contribution for the electron conduction comes from the Fe 3$d$ and Ru 4$d$ orbitals partially mediated by the As 4$p$ orbitals. These results are in good agreement with the conclusions from neutron diffraction in this Rapid Communication and previous first-principles calculations~\cite{WeiLi}.

Ascribing to the localized behaviours of Eu 4$f$ orbitals in Eu(Fe$_{0.75}$Ru$_{0.25}$)$_{2}$As$_{2}$, we establish an effective localized three-dimensional anisotropic Heisenberg model with consideration of the next-nearest neighboring magnetic exchange coupling in Eu layers for discussing magnetism in the Eu 4$f$ orbitals,
\begin{equation}
\hat{H} = J_1\sum_{\langle i,j\rangle}\vec{S}_i\vec{S}_j+J_{2}\sum_{\langle\langle i,j\rangle\rangle}\vec{S}_i\vec{S}_j+J_{\perp}\sum_{\langle i,j\rangle}\vec{S}_i\vec{S}_j,
\label{Eq1}
\end{equation}
where $\vec{S}$ is the magnitude of Eu spin. The $\langle i,j\rangle$ and $\langle\langle i,j\rangle\rangle$ denote the summation over the nearest neighbor and next-nearest neighbor sites, respectively. The parameters $J_1$ and $J_2$ describe the nearest neighboring and next-nearest neighboring intralayer exchange interactions, respectively, and $J_{\perp}$ denotes the nearest neighboring interlayer exchange interaction. From the calculated energy data for various magnetic configurations~\cite{HJXiang,SM}, the magnetic exchange couplings  $J_1=-4.10$ meV, $J_2 = 0.51$ meV, and $J_{\perp}=-0.49$ meV were found for Eu(Fe$_{0.75}$Ru$_{0.25}$)$_{2}$As$_{2}$, demonstrating the ferromagnetic ground state is consistent with the results from neutron diffraction. Comparing with the calculated magnetic exchange couplings of the parent compound EuFe$_2$As$_2$~\cite{WeiLi}, we note that the values of magnetic exchange coupling are enhanced by about five times, which stems from the extended Ru 4$d$ orbitals doping as expected in the aforementioned discussions. As a result, a three-dimensional quantum Heisenberg model with a strong anisotropy and long-range magnetic exchange coupling is suitable for describing the magnetic behaviours in the ferromagnetic superconductor Eu(Fe$_{0.75}$Ru$_{0.25}$)$_{2}$As$_{2}$.

%\section{Quantum Monte Carlo Simulations}\label{Sec4}

{\it Quantum Monte Carlo simulations.---}Based on the localized strongly anisotropic Heisenberg model, for which the model Hamiltonian is shown in Eq. (\ref{Eq1}), we have carried out a quantum Monte Carlo simulation to evaluate the universal critical exponent $\beta$ for the ferromagnetic superconductor Eu(Fe$_{0.75}$Ru$_{0.25}$)$_{2}$As$_{2}$. Although the Eu$^{2+}$ has a large magnitude of spin-7/2, previous model calculations have demonstrated that the universal critical exponent $\beta$ is irrelevant to the magnitude of large spin~\cite{hallberg_1996,jensen_1996}. It motivates us to alternatively use spin-1/2 for studying the universal critical exponent for simplicity based on the stochastic series expansion (SSE) algorithm~\cite{sandvik_1999}.

In the SSE method~\cite{SM,sandvik_1999}, the exponential operator in the partition function $Z=\mathrm{tr}\,e^{\beta'\hat{H}}$ is taken by a Taylor expansion and the trace is described by the sum over a complete set of states in a complete basis, $Z=\sum_\alpha\sum_{n=0}^{\infty}\frac{\beta'^n}{n!}\langle\alpha|(-\mathcal{H})^n|\alpha\rangle$, where $|\alpha\rangle$ is a randomly selected state and $\beta'=1/k_B T$. $k_B$ is the Boltzmann constant and $T$ is the temperature. The Hamiltonian is then rewritten as the summation of a set of operators whose matrix elements are conveniently obtained. We stochastically pick configurations from this infinite summation by means of importance sampling and average over the observable states $|\alpha\rangle$. Due to the presence of weak antiferromagnetic coupling strength $J_2$ in the anisotropic Heisenberg model in Eq. (\ref{Eq1}), it gives rise to the negative weights in the samplings. Fortunately, $J_2$ has one magnitude order smaller than that of $J_1$, the negative sign problem can be easily overcome by introducing the absolute value of weight $|W_i|$ in the calculated expectation value of the observables $\mathcal{O}$, %using the absolute value of the sampling weight.
%this problem can be dealt simply by the following technique: the expectation value of the observables can be written as
\begin{equation}
    \langle\mathcal{O}\rangle=\frac{\sum_iW_i\mathcal{O}_i}{\sum _iW_i}=\frac{\sum_iW_i\mathcal{O}_i/\sum_i|W_i|}{\sum_iW_i/\sum_i|W_i|}=\frac{\langle\mathcal{O}\:\mathrm{sgn}\:W\rangle'}{\langle\mathrm{sgn}\:W\rangle'}
\label{Eq2}
\end{equation}
where $W_i$ is the weights of the samplings and %. %Because some of the $W_i$'s might be negative, the absolute value of weight $|W_i|$ We replace the negative weights by its absolute value $|W_i|$,
$\langle\cdot\rangle'$ is denoted as the expectation value measured in these new weights. % are denoted by $\langle\cdot\rangle'$, and
Therefore, the expectation value of the observables $\mathcal{O}$ and the sign of the weights are evaluated simultaneously. %Their quotient is the expectation in the original weights.
In the numerical calculations, the three-dimensional $L\times L\times L$ sizes with $L$ ranging from 8 to 14 are performed. We set the number of bins to $50$, where the first $10$ are used for reaching the thermodynamic balance and the rest are used for measuring the physical observable quantities. Each bin contains $1000$ Monte Carlo steps~\cite{SM}.

\begin{figure}
    \centering
    \includegraphics[width=.9\linewidth]{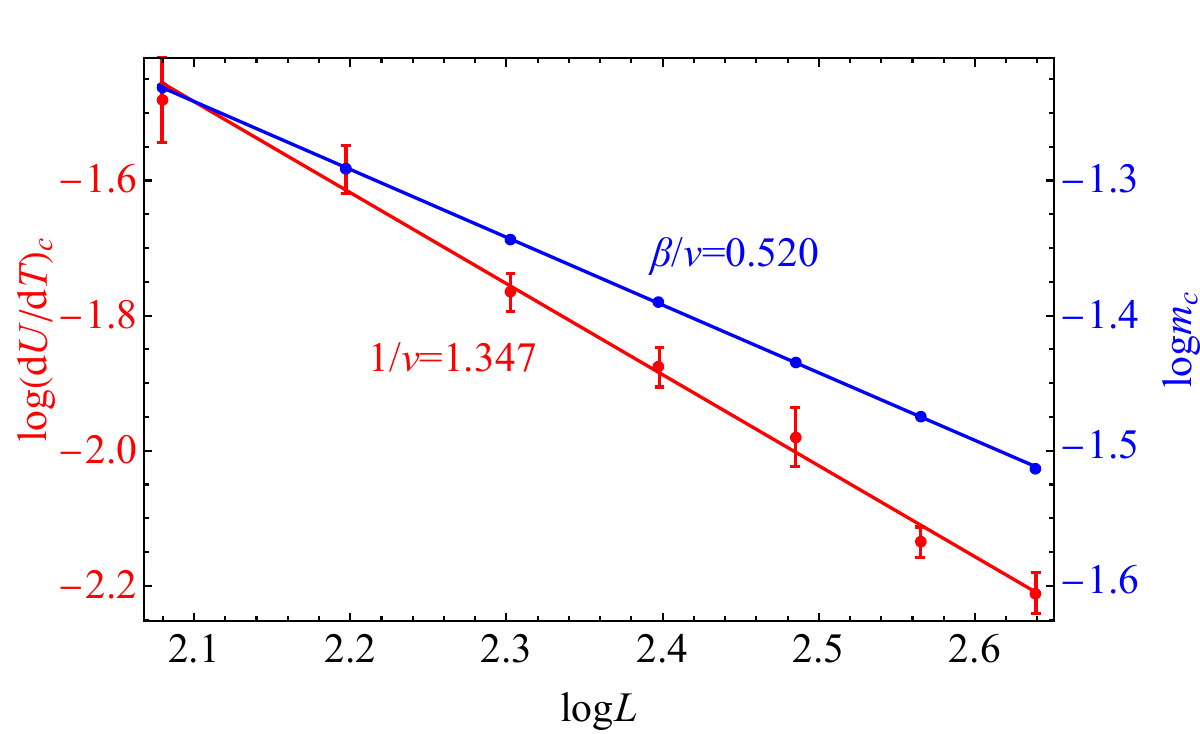}
    \caption{Log-log plot of size dependences of the magnetization $m_c$ and Binder cumulant $\mathrm{d}U/\mathrm{d}T|_c$ for the strong anisotropic three-dimensional Heisenberg model of the ferromagnetic superconductor Eu(Fe$_{0.75}$Ru$_{0.25}$)$_{2}$As$_{2}$.}
\label{fig4}
\end{figure}

The physical observable magnetization $m=\langle\sum_i\hat{S}_i^z\rangle$ and its reduced four order Binder cumulant $U=3(1-\frac{1}{2}\frac{\langle m^4\rangle}{\langle m^2\rangle^2})$ for various system sizes $L$ are evaluated numerically~\cite{SM,KChen}. Here it should be noted that the Binder cumulant reaches 1 at the paramagnetic phase whereas it reaches 0 at the ferromagnetic ordering phase. Applying the finite size scaling~\cite{SM,KChen,privman_1990}, the physical quantities follow the relations of $m_L(T)=L^{-\beta/\nu}\bar{m}(L^{1/\nu}\tau)$ and $U_L(T)=\bar{U}(L^{1/\nu}\tau)$ in the vicinity of the critical point of temperature $T_c$, where $\tau=(T-T_c)/T_c$ and $\bar{U}$ and $\bar{m}$ are universal functions that are independent of the size scale of $L$. The parameter $\nu$ is also a critical exponent, which describes the behavior of the correlation length in the vicinity of the critical temperature. At the critical temperature $T_c$, the magnetization is $m_c=m_L(T_c)=L^{-\beta/\nu}\bar{m}(0)$ $=L^{-\beta/\nu}m^*$, which is shown in Fig.~\ref{fig4}. Considering the slope of Binder culuant $\mathrm{d}U/\mathrm{d}T|_c$ as a function of size $L$, $\mathrm{d}U/\mathrm{d}T|_c=L^{1/\nu}\bar{U}'(0)$~\cite{SM}, shown in Fig.~\ref{fig4}, we finally obtain the universal critical exponent $\beta=0.386$, which is in excellent agreement with the experimental result of 0.385(13). These results obtained in the neutron diffraction experiment, the first-principles calculations, and the model simulations clearly demonstrate that the critical behavior of Eu magnetism in the ferromagnetic superconductor Eu(Fe$_{0.75}$Ru$_{0.25}$)$_{2}$As$_{2}$ belongs to the universal class of a three-dimensional anisotropic quantum Heisenberg model with long-range magnetic exchange coupling.

%\section{Discussion and Summary}\label{final}

{\it Discussion and Summary.---}By systematically combining the neutron diffraction experiment, the first-principles calculations, and the quantum Monte Carlo simulations, we have obtained a perfectly consistent universal critical exponent value of $\beta=0.385(13)$ experimentally and theoretically for Eu(Fe$_{0.75}$Ru$_{0.25}$)$_{2}$As$_{2}$. The magnetism in the Eu-based ferromagnetic superconductor is thus identified as the universal class of three-dimensional quantum Heisenberg model with a strong anisotropy and long-range magnetic exchange coupling. This study not only clarifies the nature of microscopic magnetic behaviours in the Eu-based ferromagnetic superconductors, but also theoretically open a new avenue for studying the universal critical behaviors associated with magnetic phase transitions.

%\bibliographystyle{bibtex}
%\bibliographystyle{apsrev} \bibliographystyle{apsrev}
%\begin{acknowledgments}
%\section{acknowledgements}

%\section*{ACKNOWLEDGEMENTS}
Z.Z. acknowledges support from the CURE (Hui-Chun Chin and Tsung-Dao Lee Chinese Undergraduate Research Endowment) (19925) and National University Student Innovation Program (19925). W.T.J. would like to acknowledge S. Mayr for assistance with the orientation of the crystal. W.L. acknowledges the start-up funding from Fudan University and the Natural Science Foundation of Shanghai (Grant No. 19ZR1402600).


\begin{thebibliography}{10}

\bibitem{Landau} L. D. Landau and E. M. Lifschitz, {\it Statistical Physics - Course of Theoretical Physics} Vol 5, (Pergamon, London, 1958).

\bibitem{Bulaevskii} L. N. Bulaevskii , A. I. Buzdin, M. L. Kulic, and S. V. Panjukov, {\it Coexistence of superconductivity and magnetism theoretical predictions and experimental results}, Advances in Physics \textbf{34}, 175 (1985).

\bibitem{HSJeevan} H. S. Jeevan, Z. Hossain, Deepa Kasinathan, H. Rosner, C. Geibel, and P. Gegenwart, {\it Electrical resistivity and specific heat of single-crystalline EuFe$_2$As$_2$: A magnetic homologue of SrFe$_2$As$_2$}, Phys. Rev. B \textbf{78}, 052502 (2008).

\bibitem{ZRen} Z. Ren, Q. Tao, S. Jiang, C. Feng, C. Wang, J. Dai, G. Cao, and Z. Xu, {\it Superconductivity induced by phosphorus doping and its coexistence with ferromagnetism in EuFe$_2$(As$_{0.7}$P$_{0.3}$)$_2$}, Phys. Rev. Lett. \textbf{102}, 137002 (2009).

\bibitem{SJiang} S. Jiang, H. Xing, G. Xuan, Z. Ren, C. Wang, Z.-A. Xu, and G. Cao, {\it Superconductivity and local-moment magnetism in Eu(Fe$_{0.89}$Co$_{0.11}$)$_2$As$_2$}, Phys. Rev. B \textbf{80}, 184514 (2009).

\bibitem{GCao} G. Cao, S. Xu, Z. Ren, S. Jiang, C. Feng, and Z. Xu, {\it Superconductivity and ferromagnetism in EuFe$_2$(As$_{1-x}$P$_x$)$_2$}, J. of Phys.: Condens. Matter \textbf{23}, 464204 (2011).

\bibitem{WTJin2017} W. T. Jin, J.-P. Sun, G. Z. Ye, Y. Xiao, Y. Su, K. Schmazl, S. Nandi, Z. Bukowski, Z. Guguchia, E. Feng, Z. Fu, and J.-G. Cheng, {\it Hydrostatic pressure effects on the static magnetism in Eu(Fe$_{0.925}$Co$_{0.075}$)$_2$As$_2$}, Sci. Rep. \textbf{7}, 3532 (2017).

\bibitem{WHJiao} W.-H. Jiao, Q. Tao, Z. Ren, Y. Liu, and G.-H. Cao, {\it Evidence of spontaneous vortex ground state in an iron-based ferromagnetic superconductor}, npj Quantum Materials \textbf{2}, 50 (2017).

\bibitem{WTJin2013} W. T. Jin, S. Nandi, Y. Xiao, Y. Su, O. Zaharko, Z. Guguchia, Z. Bukowski, S. Price, W. H. Jiao, G. H. Cao, and Th. Br\"{u}ckel, {\it Magnetic structure of superconducting Eu(Fe$_{0.82}$Co$_{0.18}$)$_2$As$_2$ as revealed by single-crystal neutron diffraction}, Phys. Rev. B \textbf{88}, 214516 (2013).

\bibitem{SNandi2014} S. Nandi, W. T. Jin, Y. Xiao, Y. Su, S. Price, W. Schmidt, K. Schmalzl, T. Chatterji, H. S. Jeevan, P. Gegenwart, and Th. Br\"{u}ckel, {\it Magnetic structure of the ${\mathrm{Eu}}^{2+}$ moments in superconducting ${\mathrm{EuFe}}_{2}{({\mathrm{As}}_{1\ensuremath{-}x}{\mathrm{P}}_{x})}_{2}$ with $x=0.19$}, Phys. Rev. B \textbf{90}, 094407 (2014).

\bibitem{WTJin2015} W. T. Jin, Wei Li, Y. Su, S. Nandi, Y. Xiao, W. H. Jiao, M. Meven, A. P. Sazonov, E. Feng, Yan Chen, C. S. Ting, G. H. Cao, and Th. Br\"{u}ckel, {\it Magnetic ground state of superconducting Eu(Fe$_{0.88}$Ir$_{0.12}$)$_2$As$_2$: A combined neutron diffraction and first-principles calculation study}, Phys. Rev. B \textbf{91}, 064506 (2015).

\bibitem{WeiLi} W. Li, J.-X. Zhu, Y. Chen, and C. S. Ting, {\it First-principles calculations of the electronic structure of iron-pnictide EuFe$_2$(As,P)$_2$ superconductors: Evidence for antiferromagnetic spin order}, Phys. Rev. B \textbf{86}, 155119 (2012).

\bibitem{Akbari} A. Akbari, P. Thalmeier, and I. Eremin, {\it Evolution of the multiband Ruderman{\textendash}Kittel{\textendash}Kasuya{\textendash}Yosida interaction: application to iron pnictides and chalcogenides}, New J. Phys. \textbf{15}, 033034 (2013).

\bibitem{JKoo} J. Koo, J. Park, S. K. Cho, K. D. Kim, S.-Y. Park, Y. H. Jeong, Y. J. Park, T. Y. Koo, K.-P. Hong, C.-H. Lee, J.-Y. Kim, B.-K. Cho, K. B. Lee, and H.-J. Kim, {\it Magnetic and structural phase transitions of EuFe$_2$As$_2$ studied via neutron and resonant X-ray scattering}, J. Phys. Soc. Jpn. \textbf{79}, 114708 (2010).

\bibitem{Jin2019} W. T. Jin, N. Qureshi, Z. Bukowski, Y. Xiao, S. Nandi, M. Babij, Z. Fu, Y. Su, and Th. Br\"{u}ckel, {\it Spiral magnetic ordering of the Eu moments in EuNi$_2$As$_2$}, Phys. Rev. B \textbf{99}, 014425 (2019).

%\bibitem{YCChen} Y. C. Chen, H. H. Chen, and F. Lee, {\it Quantum Monte Carlo study of the spin-1/2 Heisenberg model}, Phys. Rev. B \textbf{43}, 11082 (1991).

\bibitem{SM} See Supplemental Materials for the the critical exponents values for several common theoretical models, the experimental details, the evaluations of the magnetic exchange coupling constants, and the details of numerical quantum Monte Carlo calculations for the critical exponents.


\bibitem{SNandi2009} S. Nandi, A. Kreyssig, Y. Lee, Yogesh Singh, J. W. Kim, D. C. Johnston, B. N. Harmon, and A. I. Goldman, {\it Magnetic ordering in EuRh$_2$As$_2$ studied by x-ray resonant magnetic scattering}, Phys. Rev. B \textbf{79}, 100407(R) (2009).

\bibitem{Kagawa} F. Kagawa, K. Miyagawa, and K. Kanoda, {\it Unconventional critical behaviour in a quasi-two-dimensional organic conductor}, Nature \textbf{436}, 534 (2005).

\bibitem{Jiao_11} W. H. Jiao, Q. Tao, J. K. Bao, Y. L. Sun, C. M. Feng, Z. A. Xu, I. Nowik, I. Feiner, and G. H. Cao, {\it Anisotropic superconductivity in Eu(Fe$_{0.75}$Ru$_{0.25}$)$_2$As$_2$ ferromagnetic superconductor}, Europhys. Lett. \textbf{95}, 67007 (2011).

\bibitem{Xiao_09} Y. Xiao, Y. Su, M. Meven, R. Mittal, C. M. N. Kumar, T. Chatterji, S. Price, J. Persson, N. Kumar, S. K. Dhar, A. Thamizhavel, and Th. Brueckel, {\it Magnetic structure of ${\text{EuFe}}_{2}{\text{As}}_{2}$ determined by single-crystal neutron diffraction}, Phys. Rev. B \textbf{80}, 174424 (2009).

\bibitem{Coppens_65} P. Coppens, L. Leiserowitz, and D. Rabinovich, {\it Calculation of absorption corrections for camera and diffractometer data}, Acta Crystallogr \textbf{18}, 1035 (1965).

\bibitem{Rodriguez-Carvajal_93} J. Rodr\'iguez-Carvajal, {\it Recent advances in magnetic structure determination by neutron powder diffraction}, Physica B \textbf{192}, 55 (1993).

\bibitem{Wills_00} A. S. Wills, {\it A new protocol for the determination of magnetic structures using simulated annealing and representational analysis (SARAh)}, Physica B \textbf{276-278}, 680 (2000).

\bibitem{LZhang} L. Zhang, J. Fan, L. Li, R. Li, L. Ling, Z. Qu, W. Tong, S. Tan, and Y. Zhang, {\it Critical properties of the 3D-Heisenberg ferromagnet CdCr$_2$Se$_4$}, Europhys. Lett. \textbf{91}, 57001 (2010).

\bibitem{DJSingh} D. J. Singh and L. Nordstrom, {\it Planewaves, Pseudopotentials, and the LAPW Method}, 2nd ed. (Springer-Verlag, Berlin, 2006), pp. 1C134.

\bibitem{PBlaha} P. Blaha, K. Schwarz, G. Madsen, D. Kvasnicka, and J. Luitz, in WIEN2K, {\it An Augmented PlaneWave + Local Orbitals Program for Calculating Crystal Properties}, edited by K. Schwarz (Technical Univievsity Wien, Austria, 2001).

\bibitem{PBE} J. P. Perdew, K. Burke, and M. Ernzerhof, {\it Generalized Gradient Approximation Made Simple}, Phys. Rev. Lett. \textbf{77}, 3865 (1996).

\bibitem{potentials} W.-C. Huang, W. Li, and X. Liu, {\it Exotic ferromagnetism in the two-dimensional quantum material C$_3$N}, Front. Phys. \textbf{13}, 137104 (2018).

\bibitem{HJXiang} H. J. Xiang, E. J. Kan, Su-Huai Wei, M.-H. Whangbo, and X. G. Gong, {\it Predicting the spin-lattice order of frustrated systems from first principles}, Phys. Rev. B \textbf{84}, 224429 (2011).

\bibitem{hallberg_1996} K. Hallberg, X. Q. G. Wang, P. Horsch, and A. Moreo, {Critical behavior of the $S=3/2$ antiferromagnetic Heisenberg chain}, Phys. Rev. Lett. \textbf{76}, 4955 (1996).

\bibitem{jensen_1996} I. Jensen, A. J. Guttmann, and I. G. Enting, {\it Low-temperature series expansions for the square lattice Ising model with spin $S>1$}, J. Phys. A: Math. Gen. \textbf{29}, 3805 (1996).

\bibitem{sandvik_1999} A. W. Sandvik, {\it Stochastic series expansion method with operator-loop update}, Phys. Rev. B \textbf{59}, R14157 (1999).

\bibitem{KChen} K. Chen, A. M. Ferrenberg, and D. P. Landau, {\it Static critical behavior of three-dimensional classical Heisenberg models: A high-resolution Monte Carlo study}, Phys. Rev. B \textbf{48}, 3249 (1993).

\bibitem{privman_1990} V. Privman, {\it Finite-size scaling and numerical simulation of statistical systems}, (Singapore: World Scientific, Singapore, 1990).

%\bibitem{MSato} M. Sato and Y. Ando, {\it Topological superconductors: A review}, Rep. Prog. Phys. \textbf{80}, 076501 (2017).

%\bibitem{Pfleiderer} C. Pfleiderer, {\it Superconducting phases of f-electron compounds}, Rev. Mod. Phys. \textbf{81}, 1551 (2009).

%\bibitem{FF} P. Fulde and R. A. Ferrell, {\it Superconductivity in a strong spin-exchange field}, Phys. Rev. \textbf{135}, A550, (1964).

%\bibitem{LO} A. I. Larkin and R. A. Ferrell, {\it Superconductivity in a strong spin-exchange field}, Phys. Rev. \textbf{135}, A550, (1964).

%\bibitem{CHolm1993} C. Holm and W. Janke, {\it Critical exponents of the classical three-dimensional Heisenberg model: A single-cluster Monte Carlo study}, Phys. Rev. B \textbf{48}, 936 (1993).

%\bibitem{Fakioglu1988} S. Fakioglu, {\it Critical exponents for two-dimensional Heisenberg ferromagnet with long-range interaction}, Il Nuovo Cimento D \textbf{10}, 1161 (1988).

%\bibitem{Ferrenberg1991} A. M. Ferrenberg and D. P. Landau, {\it Critical behavior of the three-dimensional Ising model: A high-resolution Monte Carlo study}, Phys. Rev. B \textbf{44}, 5081 (1991).

%\bibitem{CNYang1952} C. N. Yang, {\it The Spontaneous Magnetization of a Two-Dimensional Ising Model}, Phys. Rev. \textbf{85}, 808 (1952).

%\bibitem{Whangbo} M.-H. Whangbo, H.-J. Koo, and D. Dai, {\it Spin exchange interactions and magnetic structures of extended magnetic solids with localized spins: theoretical descriptions on formal, quantitative and qualitative levels}, Journal of Solid State Chemistry \textbf{176}, 417 (2003).

%\bibitem{Sandvik2010} A. W. Sandvik, {\it Computational Studies of Quantum Spin Systems}, AIP Conf. Proc. \textbf{1297}, 135 (2010).

\end{thebibliography}
\end{document}